\documentclass[aps,pra,reprint,showpacs,superscriptaddress]{revtex4-2} 
\usepackage{graphicx}
\usepackage{amsmath,mathtools,verbatim}
\usepackage{amssymb}
\usepackage{mathtools}
\usepackage{amsmath}
\usepackage{bm} 
\usepackage{epstopdf}
\usepackage{braket}
\usepackage{siunitx}
\usepackage[version=4]{mhchem}
\usepackage[colorlinks=true, citecolor=blue, urlcolor=blue ]{hyperref}
\usepackage{xcolor}
\usepackage{hyperref}
\usepackage{xspace} 
\usepackage{import}

\newcommand{\hide}[1]{}

\newcommand{\be}{\begin{equation}}
\newcommand{\ee}{\end{equation}}
\newcommand{\bqa}{\begin{eqnarray}}
\newcommand{\eqa}{\end{eqnarray}}

\newcommand{\IITBEE}{Department of Electrical Engineering, Indian Institute of Technology Bombay, Mumbai 400076, India}
\newcommand{\IITBPhys}{Department of Physics, Indian Institute of Technology Bombay, Mumbai 400076, India}
\newcommand{\QuICST}{Centre of Excellence in Quantum Information, Computation, Science and Technology, Indian Institute of Technology Bombay, Mumbai 400076, India}
\newcommand{\IITK}{Department of Physics, Indian Institute of Technology Kanpur, UP-208016, India}

\begin{document}

\title{Cavity-Driven Multispectral Gain for High-Sensitivity NV Center Magnetometers}

\author{Himanshu Kumar} 
\thanks{These authors contributed equally to this work.}
\affiliation{\IITBEE}
\author{Rahul Gupta}
\thanks{These authors contributed equally to this work.}
\affiliation{\IITBPhys} 
\author{Saikat Ghosh} \affiliation{\IITK}
\author{Himadri Shekhar Dhar} \affiliation{\IITBPhys} \affiliation{\QuICST} 
\author{Kasturi Saha} \affiliation{\IITBEE} \affiliation{\QuICST} 

\date{\today}    

\begin{abstract}
We report a cavity-enabled solid-state magnetometer based on an NV ensemble coupled with a dielectric cavity, achieving 12 pT/$\sqrt{\text{Hz}}$ sensitivity and a nearly threefold {gain} from multispectral features. The features originate from cavity-induced splitting of the NV hyperfine levels and {leverages} robust quantum coherence in the doubly dressed states of the system {to achieve high sensitivity}. We project simulated near-term sensitivities approaching 100 fT/$\sqrt{\text{Hz}}$, close to the Johnson–Nyquist limit. Our results establish frequency multiplexing as a new operational paradigm, offering a robust and scalable quantum resource for metrology under ambient conditions.
\end{abstract}

\maketitle

\section{Introduction}
Quantum resources such as entanglement~\cite{Horodecki2009} play a key role in quantum metrology, especially in achieving super-enhanced sensitivities~\cite{Giovannetti2004,Boixo2008}. However, the generation of high-fidelity entanglement in solid state quantum sensors
typically involves complex control schemes and is extremely fragile -- for instance, the enhanced Heisenberg scaling due to entanglement may disappear in the presence of even infinitesimal environmental noise~\cite{Dobrzanski2012}.
Alternatively, for sensors operating in ambient conditions, superposition of quantum states or quantum coherence~\cite{Streltsov2017} is a more robust and cost-effective resource that can be scaled deterministically to yield significant metrological advantage~\cite{Tilma2010,Zhang2019}.

Nitrogen-vacancy (NV) centers in diamond present an ideal platform for a coherence based approach ~\cite{Taylor2008,JohnFBarry.pnas,PhysRevB.80.115202}.
While they exhibit exceptionally long coherence times~\cite{PhysRevB.85.115303,PhysRevB.80.041201} and operate at room temperatures~\cite{PhysRevLett.108.197601,PhysRevLett.114.145502}, their sensitivity has historically been limited due to low photon collection efficiency of optical readout. Recent hybrid architectures that couple an ensemble of NV spins with a microwave (MW) cavity have overcome these limitations by exploiting the strong enhancement due to cavity quantum electrodynamics~\cite{breeze2018continuous,day2024room,Eisenach2021cavity}.
However, an alternative regime exists that relates to strongly-driven but weakly interacting spin-cavity system, which can serve as a resource for enhanced sensitivity measurements. This regime is relatively less explored. 
In this configuration, strong classical fields drive the spin ensemble into a high-coherence subspace manifested via Autler-Townes splitting (ATS)~\cite{Autler1955} and formation of Mollow triplets~\cite{Mollow1969}, while the cavity serves as a broadband, low-latency transducer for readout.

In this work, we experimentally demonstrate a quantum-enhanced magnetometer that leverages the rich coherence structure of ``doubly dressed" states in a solid-state cavity QED system. By strongly driving an ensemble of NV spins, we generate Autler-Townes splitting for each hyperfine transition, creating a multispectral manifold of nine distinct resonant peaks. This cascaded interaction, where spins are first dressed by the strong drive and then weakly coupled to the cavity vacuum, allows us to treat the cumulative coherence of these modes as a metrological resource. We show that simultaneous homodyne detection across this multispectral landscape yields a sensitivity enhancement factor of 3, achieving a magnetic field noise floor of \SI{12}{\pico\tesla\per\sqrt{\hertz}} at room temperature. Our findings are supported by a full quantum model treating the system as a cascaded Tavis-Cummings ensemble in the dispersive limit, predicting that this architecture can reach sensitivities approaching \SI{100}{\femto\tesla\per\sqrt{\hertz}}. 
{Moreover, our cascaded interaction model can be extended to other hybrid platforms, including multimode and polarization-engineered cavities, which can lead to a scalable framework towards achieving sensitivity beyond the single mode standard quantum limit.} 

\begin{figure*}[htbp] 
\center
\includegraphics[width=\textwidth]{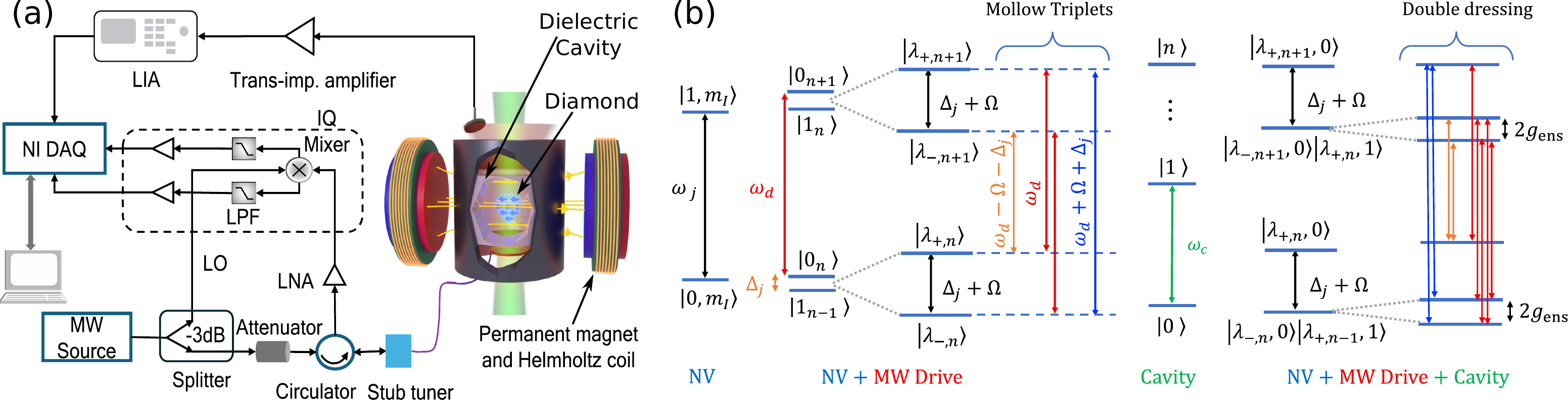} 
\caption{(a) Schematic of the experimental setup showing a diamond inside a dielectric cavity, housed inside an Aluminum shell and placed between magnets. Other components include low-noise amplifier (LNA), local oscillator (LO), low-pass filter (LPF), lock-in amplifier (LIA) and National Instruments data acquisition (NI DAQ) system. 
(b) Energy level diagram of an NV spin state, with frequency $\omega_s = \omega_j$ and nuclear spin $m_I$, dressed with a strong classical field, with frequency $\omega_d$. The dressed states are $\ket{\lambda_{\pm,n}}$ and the transitions clearly show the Mollow triplets. 
Here, $\Delta_j = \omega_j - \omega_d$ and $\Omega$ is the driving strength or the Rabi frequency. The double dressing occurs when the spin states hybridize with a cavity photon of frequency $\omega_c$,  leading to states $\ket{\lambda_{-,n},0} \pm \ket{\lambda_{+,n-1},1}$. This leads to additional splitting and new transition lines, governed by the spin-cavity coupling $g_\text{ens}$. 
}
\label{fig1}
\end{figure*}

\section{Main results}

\begin{figure}[t] 
\center
\includegraphics[width=0.48\textwidth]{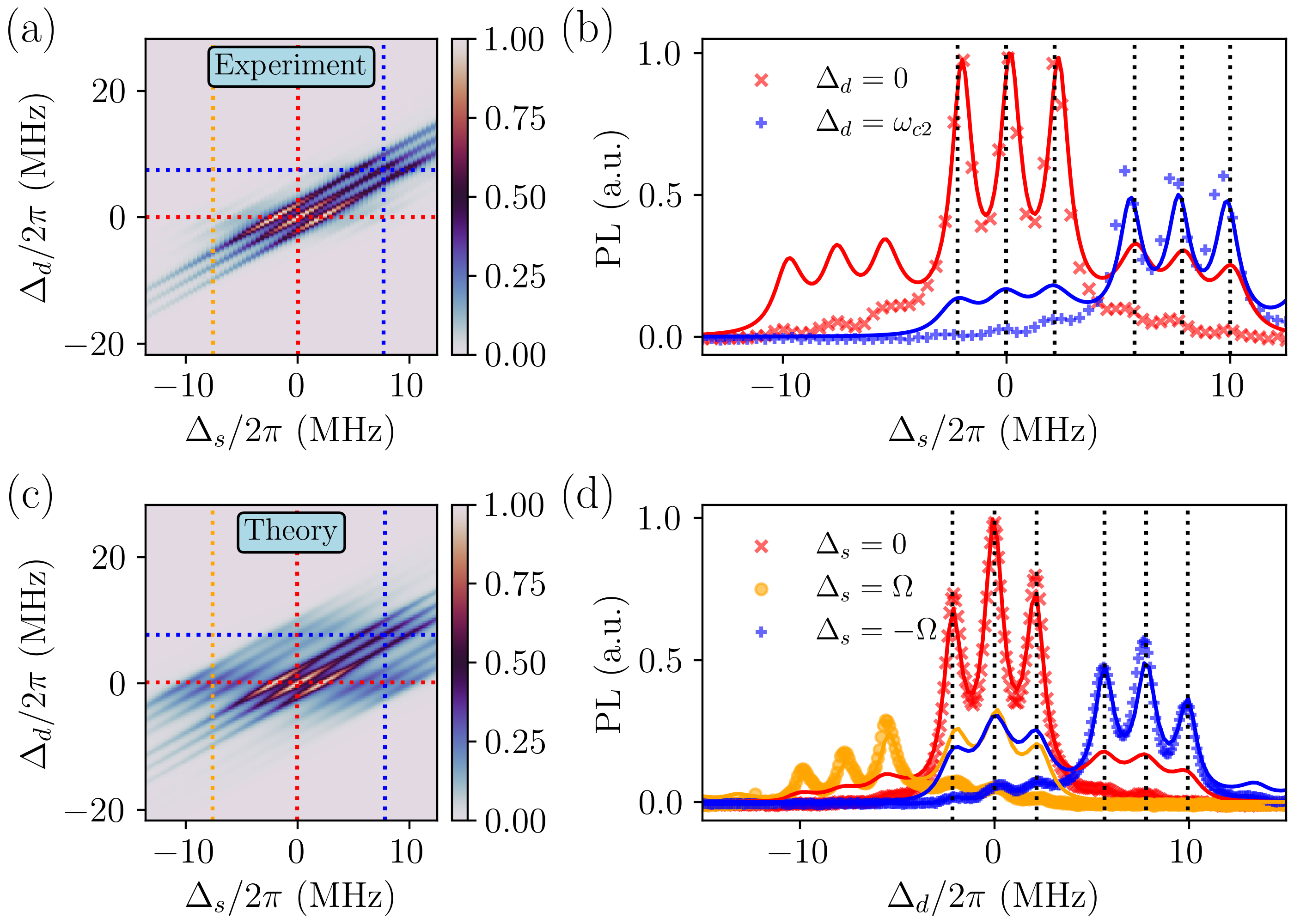} 
\caption{Amplitude modulated ODMR spectrum. Measurements at \SI{1.5}{\kilo\hertz} modulation frequency, \SI{100}{\milli\s} lock-in time constant, and \SI{-45}{dBm} power. The plots in (a,c) show normalized experimental and simulated ODMR spectra, respectively, for sweeping spin and drive frequency detunings, $\Delta_s$ and $\Delta_c$. Plot (b) shows the spectra for values of $\Delta_s$ along the two horizontal lines in (a), represented by red-cross and blue-plus symbols in the scatter plot. The solid lines correspond to the simulated spectra for the same values. The scatter plot in (d) shows the same, but for three vertical lines in (a), represented by red-cross, yellow-circle and blue-plus symbols.}
\label{fig2}
\end{figure}

In our experiment, a high density of ensemble NV centers in a bulk diamond are placed inside a dielectric resonator, which is driven by a strong microwave field, as illustrated in Fig.~\ref{fig1}(a). 
The energy levels of the NV center are described in terms of the electron spin $m_s$ and the \ce{ ^14N} nuclear spin $m_I$ numbers, where $m_s,m_I\in{\{0,\pm 1\}}$. The microwave drive is close to resonance with the electronic transition between $m_s=1$ and $m_s=0$ levels.  
In strong driving regime, the two-levels are dressed by the electromagentic field such that $\ket{\lambda_{\pm,n}}\rightarrow\ket{0_{n+1},m_I}\pm\ket{1_{n},m_I}$, for each of the hyperfine nuclear levels corresponding to $m_I$. Here, $n$ is the photonic excitation in the system. The dressed states are split by the Rabi frequency $\Omega$ -- this effect is known as Autler-Townes splitting~\cite{Autler1955}. Transitions between the $\ket{\lambda_{\pm,n+1}}$ and $\ket{\lambda_{\pm,n}}$ dressed states leads to three fluorescence peaks in the spectrum -- called the Mollow triplets~\cite{Mollow1969}. 
Figure~\ref{fig1}(b) sketches the dressing and splitting of the different energy levels in an NV center under the ATS effect. Importantly, the Mollow triplets are observed for each of the nuclear spins $m_I$, thus giving rise to nine peaks in the spectrum, as observed in the optically detected magnetic resonance (ODMR) measurements shown in Fig.~\ref{fig2}.
For optical measurements, the bias field is aligned such that all four $\mathrm{\ce{NV^-}}$ classes experience the same field and their spectrum overlaps. 
Figure~\ref{fig2}(a) shows the experimental ODMR spectrum in terms of detuning of the spin {$\Delta_s = \omega_s-\omega_c$ and the drive $\Delta_d=\omega_d-\omega_c$, where $\omega_s$, $\omega_d$ and $\omega_c$ are the spin, drive and cavity frequencies, respectively. The detunings} are controlled by changing the current in the Helmholtz coil and sweeping the microwave frequency. At resonance, $\Delta_s = \Delta_d = 0$, the amplitude of the three hyperfine transitions increases due to the stronger driving enabled by the cavity mode. Each hyperfine transition, gives rise to a Mollow triplet separated by the Rabi frequency, determined by the drive strength $\Omega$. The measurements show good agreement with the theoretical calculation of the spectrum, as highlighted in Fig.~\ref{fig2}(c). The horizontal and vertical line cuts are further shown in Fig.~\ref{fig2}(b) and Fig.~\ref{fig2}(d) respectively for clearly demonstrating the evidence of Mollow triplets. The scatter plots correspond to experimental data while the solid lines refer to the simulated data at the same $\Delta_s$ and $\Delta_d$.

\begin{figure}[t] 
\center
\includegraphics[width=0.48\textwidth]{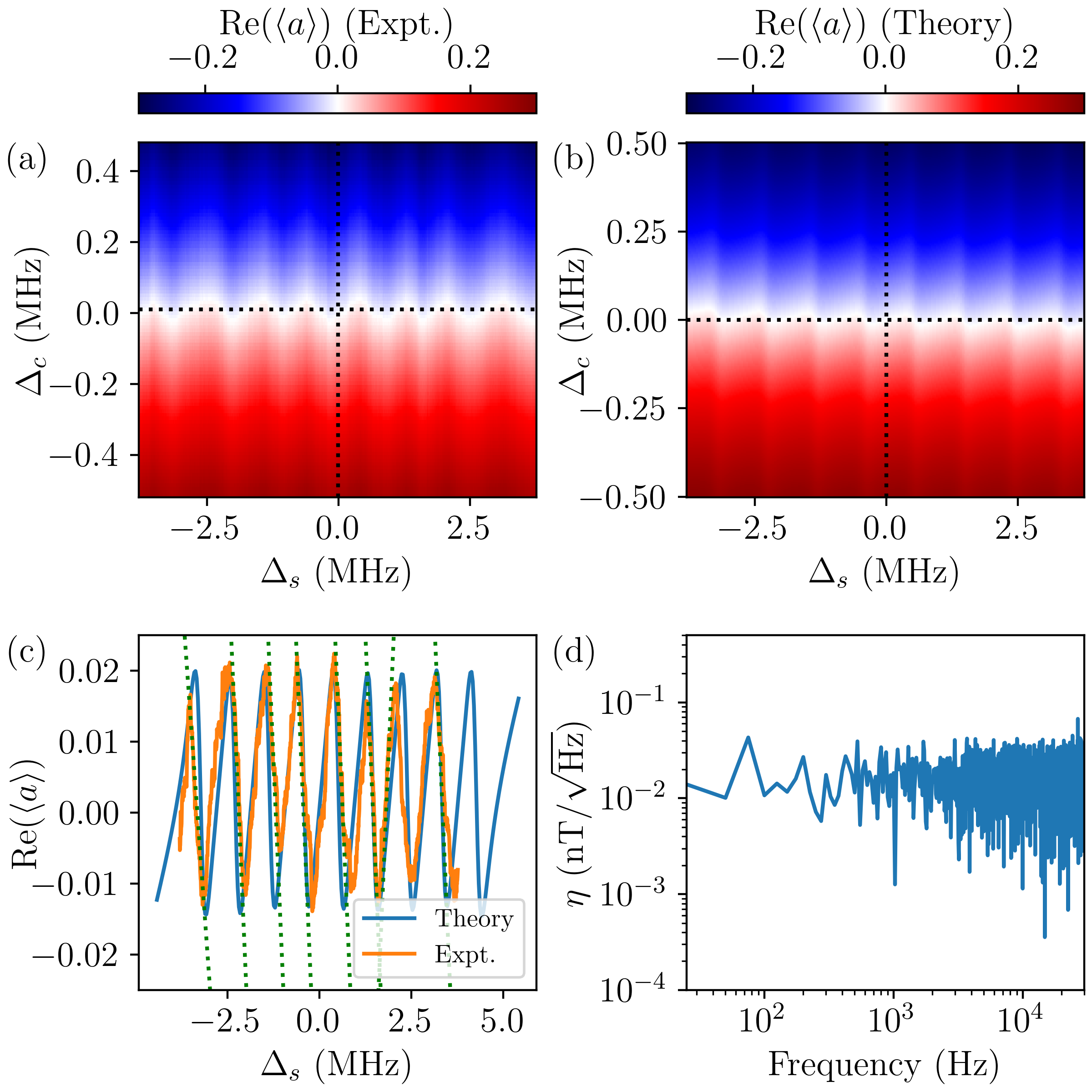} 
\caption{Homodyne measurements. Plots (a) and (b) show experimentally measured averaged reflected quadrature spectrum and simulated values, respectively. Experimental measurements were done for \SI{0}{dBm} LO power, \SI{-40}{dBm} excitation power, and \SI{700}{\kilo\hertz} LPF bandwidth. The spin and drive detuning $\Delta_d$ and $\Delta_s$, with respect to cavity frequency, are adjusted in steps of \SI{10}{\kilo\hertz} and \SI{5.2}{\kilo\hertz}.
Plot (c) shows the quadrature spectrum along the horizontal cut lines for experimental (orange) and simulation (blue) data. The green dotted straight lines 
are used to calculate spectral slope.
Plot (d) shows magnetically equivalent noise density spectrum showing an experimental sensitivity of \SI{12}{\pico\tesla\per\sqrt{\hertz}}.}
\label{fig3}
\end{figure}

The second cavity-enabled quantum effect arises from the coherent interaction between the spin ensemble and the cavity photons, resulting in a vacuum Rabi splitting that scales as {$g_\text{ens}=g\sqrt{N_s}$, where $g$ is the single spin-cavity coupling and $N_s$ is the number of spins in the ensemble.}
{Depending on cavity resonance, the spin states $\ket{\lambda_{-,n+1}}$ and
$\ket{\lambda_{+,n}}$ can hybridize with 
the Fock state of the photon to form $\ket{\lambda_{-,n+1},0}\pm \ket{\lambda_{+,n},1}$, as shown in Fig.~\ref{fig1}(b). This causes the side ATS peaks to split into two, while the central peak splits into three. Note that if $g_\text{ens}$ is small, the splitting of peaks can not be resolved.
However, analyzing the microwave cavity reflected spectrum in Figs.~\ref{fig3}(a)-(b) shows distinct avoided crossings due to the coherent Rabi splitting in each of the peaks corresponding to the Mollow triplets of the NV spins.}
The microwave spectrum is measured using a vector network analyzer and homodyne measurements.  
The measured reflected spectrum for the in-phase quadrature, is shown in Fig.~\ref{fig3}(a), as a function of the spin and cavity detuning, $\Delta_s$ and $\Delta_c$, which are varied by sweeping the current in the Helmholtz coils and the microwave probe frequency.
Figure~\ref{fig3}(b) shows the reflected phase quadrature as calculated from our theoretical model, which shows excellent agreement with the observed effects from ATS and the energy splitting or avoided crossing due to spin-photon dressing. 

\section{Sensitivity measurements}
The reflected microwave readout serves as a highly sensitive probe for detecting magnetic field variations, with the goal of leveraging quantum effects {such as coherence} enabled by cavity-enhanced coupling. In particular, the emergence of additional spectral peaks associated with {the doubly dressed states}
can be harnessed to achieve a substantial sensitivity gain. The sensitivity can be estimated using the relation,
\begin{equation}
    \eta_{\rm{exp}}=\frac{1}{\sqrt{\sum_{j=1}^M \eta^{-2}_j}},~~~\eta_{j} = \frac{\sqrt{3}\sigma_j\sqrt{\tau}}{\gamma_e\times S_j},
\label{eq:sense}
\end{equation}
where $M$ is the number of spectral peaks observed and $\eta_{j}$ is the sensitivity measured for each spectral peak.
The $\sqrt{3}$ factor accounts for the equal projection along all \ce{NV} classes
and $\sigma_j$ is the standard deviation of the time-trace for $j^{\rm{th}}$ peak with integration time $\tau$. 
The gyromagnetic ratio for electron spin is given by $\gamma_e$.
One of the key quantities is $S_j=|dV_{Q}/d\omega_s|_{\rm{max}}^{(j)}$, which is the maximum slope of the quadrature signal at $j^{\rm{th}}$ peak with spin $\omega_j$. This captures the response of the signal to change in the spin transition frequency and is obtained by linear fitting the quadrature spectrum at fixed cavity detuning corresponding to maximally sensitive point. 

The experimental results shown in Fig.~\ref{fig3}, exhibit 8 of the 9 spectral peaks arising from the Autler Townes effect, within the range of sweeping the transition frequency. {The coherent, dressed states and} the Mollow triplets arising from each of the hyperfine levels of the \ce{NV} spin, perfectly fit the theoretical simulations, and gives us a close to 3-fold or 9~dB of enhancement over single peak measurements.  
The sensitivity is obtained using the slope of the dotted lines in Fig.~\ref{fig3}(c), with Fig.~\ref{fig3}(d) showing the experimentally measured sensitivity of \SI{12}{\pico\tesla\per\sqrt{\hertz}}. Table~\ref{tab:table1} presents a comparison between the sensitivity achieved in this work and that reported in other cavity-enhanced, microwave readout experiments. We find that our performance is competitive, particularly in light of the low laser power used for spin polarization.
{However, the cavity driven multispectral gain opens up a new operational paradigm, with potential for significant improvement in sensitivity.}

\begin{table*}[t]
\caption{\label{tab:table1}%
Comparison of sensitivity in cavity enabled \ce{NV^-} ensemble experiments reported using non-optical readout}
\begin{ruledtabular}
\begin{center}
\begin{tabular}{l c c}
\textrm{Reference}&
\multicolumn{1}{c}{\textrm{Sensitivity}}&
\multicolumn{1}{c}{\textrm{Parameters}}\\
\colrule
Cavity enabled sensing & \SI{3}{\pico\tesla\per\sqrt\hertz} & Power = 8 W \\
Eisenach et al.~\cite{Eisenach2021cavity}&& Quantum-grade CVD diamond, $\ce{NV^-}=5.0~\textrm{ppm}$\\
&&\\
cQED with optimal cooling& \SI{576}{\femto\tesla\per\sqrt\hertz} & Power = 12 W \\
Wang et al.~\cite{wang2024spin}&&\ce{^12C} \textrm{enriched}, $\ce{NV^-}=4.0~\textrm{ppm}$\\
&& \\
\textbf{Present work} & \SI{12}{\pico\tesla\per\sqrt\hertz} and \SI{28}{\pico\tesla\per\sqrt\hertz} & Power = 2.1 W \\
&(multiple and single peak)& Quantum-grade CVD diamond, $\ce{NV^-}=4.5~\textrm{ppm}$\\
\end{tabular}
\end{center}
\end{ruledtabular}
\end{table*}

\section{Theoretical Model}
To model the coupling of the \ce{NV} centers with a microwave cavity, we use the quantum Tavis-Cummings model~\cite{Tavis1968,Shore1993}. In the absence of nuclear spin driving 
to induce transitions between different hyperfine manifolds, 
each nuclear spin projection $m_I$ effectively behaves as an independent two-level system, with transitions occurring between the $m_s = 0$ and $m_s =1$ electronic spin states. Hence, we can model the system using three spins corresponding to 
$m_I\in\{-1,0,+1\}$~\cite{Kirova2017} simultaneously coupled with the microwave cavity resonator with frequency $\omega_c$. To study the Autler Townes effect, the cavity is driven by a strong microwave field with a large photon number $n$. In the linear regime, the cavity operator $a\rightarrow \sqrt{n}+\delta a$ and the dynamics is primarily restricted to the collective bright modes. 
The effective Hamiltonian is
\begin{align}
    H &= \Delta_c a^\dag a + \sum_{j=-1}^1 \left[\frac{\Delta_j}{2} \sigma_j^z + g_{\rm{ens}}(a \sigma_j^+ + a^\dag \sigma_j^-)\right]\nonumber\\
    &+\frac{\Omega}{2}\left(\sigma_j^+ + \sigma_j^-\right),~~
    \label{DoubleDressingHamiltonian}
\end{align}
where $a = \delta a$ is the cavity fluctuation operator and $\sigma_j's$ are the Pauli spin operator, respectively, for each hyperfine spin $j=m_I$, with transition frequency $\omega_j$. Here, $\Delta_c=\omega_c-\omega_d$ and $\Delta_j=\omega_j-\omega_d$ are the cavity and spin detuning with respect to the drive field. 
The effective coupling of spins with the cavity is 
$g_{\rm{ens}}$, with $\Omega = 2g_{\rm{ens}}\sqrt{n}$ is driving strength and the Rabi frequency. To account for dissipation and losses in the system, we consider a Lindblad master equation~\cite{Breuer2002}
\begin{align}
    &\dot{\rho}=\frac{1}{i}\left[H,\rho\right] - \frac{\gamma_0}{2}\mathcal{D}\left[a\right]\rho - \sum_{j,k}\left[\frac{\gamma_{k}}{2}\mathcal{D}[\sigma_j^k]\rho\right],
\label{eq:master_eq}
\end{align}
where $\mathcal{D}\left[x\right]\rho=\{x^\dag x,\rho\}-2x\rho x^\dag$, where $x\in\{a,\sigma_j^-,\sigma_j^z\}$. Moreover, $\sigma^k_j = \{\sigma^-_j,\sigma^z_j\}$ and $\gamma_k = \{\gamma_{\|},\gamma_{\perp}\}$. 
The cavity loss rate is $\gamma_0$, the spin relaxation is $\gamma_{\|}$ and spin dephasing $\gamma_{\perp}$, arising from the spin cooling and the inhomogeneity of the bulk NV ensemble.
The spectrum is obtained by numerically calculating the Fourier transform of the two-time correlation function of the output field 
\begin{align}
    S(\omega)=\int_{-\infty}^{\infty}\langle a^\dag (\tau)a(0)\rangle d\tau, 
\end{align}
which results in excellent agreement with the experimental ODMR data shown in Fig.~\ref{fig2}(a). Note that the model perfectly predicts the ATS and Mollow triplets arising from the coherent dressed states for each of the nuclear spin in the system. {Importantly, the spacing between the different peaks are given by Rabi frequency $\Omega$ and the detuning between the three spins correspond to the spacing between the different nuclear spins.}

To model the non-optical reflected microwave quadrature as measured by the homodyne detection, we focus on the spin-photon dressing that occurs close to each of the Mollow peaks. Note that experimental results show cavity enabled energy splitting and avoided crossing near the spectral peaks, {which correspond to the coupling of the cavity fluctuations with the spins. These occur above the average photons in the cavity that drives the ATS splitting.} 
For simplicity, we use an analytical model, where we treat each of the dressed states as effective two-level systems with resonant frequency $\omega_j=k\Omega + l\Delta$, where $(k,l)\in\{-1,0,1\}$ are indices for the ATS peaks with spacing $\Omega$ and for the hyperfine levels with spacing $\Delta$, respectively. 
Each of these are treated as an independent TLS spin mode coupled to the same cavity field, 
with the Hamiltonian given by
\begin{equation}
    \mathcal{H}=\Delta_c a^{\dag} a + \sum_{j}\Delta_{j} \sigma^+_{j}\sigma^-_{j} +  g_{\rm{ens}}\sum_{j}\left(a^\dag \sigma^-_{j} + a\sigma^+_{j}\right),
\end{equation}
Note that the microwave drive is treated as a probe field to study the energy splitting or avoided crossing arising due to the strong coupling. The open dynamics of the system can now been modeled using quantum Langevin equations~\cite{Gardiner1985}, given by 
\begin{align}
    \dot{a}&=-\left(i\frac{\gamma_0}{2} + \Delta_c\right)a -\sum_{j}ig\sigma^-_j + \sqrt{\gamma_0}a^{\rm{mw}}+\sqrt{\gamma_0}a^{\rm{in}},\nonumber\\
    \dot{\sigma}^-_j&=-\left(i\frac{\Gamma}{2} + \Delta_j\right)\sigma^-_j +igP\sigma^-_j + \sqrt{\Gamma}b_j^{\rm{in}}.
\label{eq:langevin}
\end{align}
Here $P=\langle\sigma_j^z\rangle\approx-1$ is the spin polarization, which is enhanced by optical cooling of the ensemble. The operators $a^{\rm{in}}$ and $b_j^{\rm{in}}$ are the input noise terms that include cavity and thermal noise. $\Gamma$ is the effective loss term that takes into account dissipation due to optical cooling and inhomogeneity in the ensemble. In steady-state, the photon amplitude is given by
\begin{align}
    \alpha&=\frac{-\sqrt{\gamma_0}a^{\rm{mw}}}{i\Delta_c +\gamma_0/2 -g^2 P \sum_j\left(i\Delta_j + \Gamma/2\right)^{-1}}.
\label{eq:ss_photon}
\end{align}
We plot $\langle Q\rangle=\sqrt{2}\rm{Re}(\alpha)$ in Fig.~\ref{fig3}(a) and find remarkable agreement with homodyne measurements. Moreover, the quadrature noise spectrum can be obtained through the fluctuations around steady states, which allows us to provide theoretical estimates for the sensitivity using Eq.~\eqref{eq:sense}. 

\begin{figure}
    \centering
    \includegraphics[width=\linewidth]{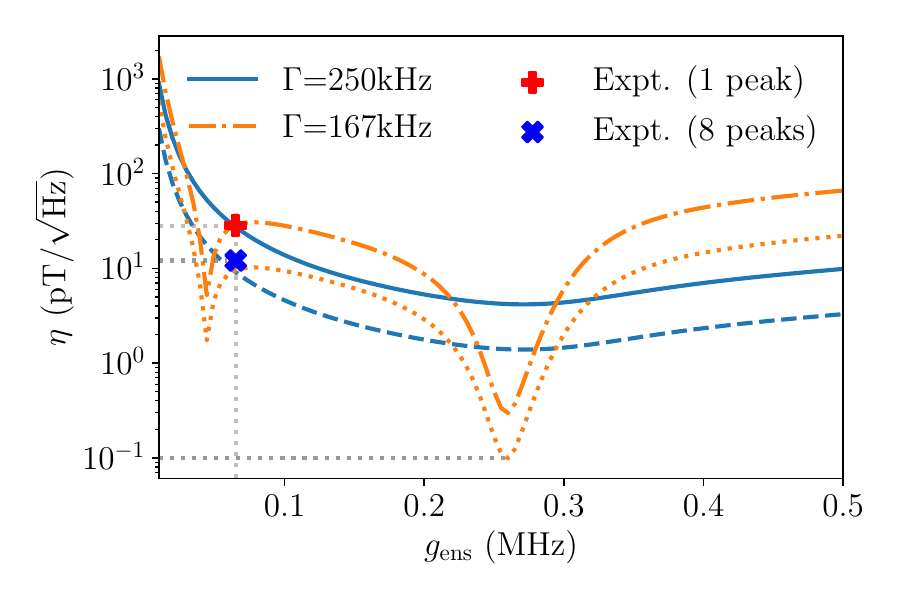}
    \caption{Theoretically projected sensitivity. The plot shows variation of sensitivity with $g_{\rm{ens}}$, for $\Gamma=250$ kHz (solid-blue) and $\Gamma=167$ kHz (orange-dash-dot). 
    The blue-dashed and orange-dotted lines show sensitivity with a multispectral 3 fold enhancement due to coherent, doubly dressed states. Experimentally measured sensitivities 28~pT/$\sqrt{\rm{Hz}}$
    (red) and 12~pT/$\sqrt{\rm{Hz}}$ (green) are for single and multiple peaks measurements. 
    Near term projection show sensitivity as high as 100~fT/$\sqrt{\rm{Hz}}$ for $g_{\rm{ens}}\approx$~0.26~MHz and $\Gamma=167$~kHz.
    }
    \label{fig:sense_comp}
\end{figure}

Figure~\ref{fig:sense_comp} presents a sensitivity analysis based on our quantum model of an NV spin ensemble coupled to a microwave cavity. 
The simulations show a $\sqrt{1/M}$ enhancement in sensitivity across different parameters, arising from $M = 9$ spectral peaks 
under the combined effects of Autler–Townes and vacuum Rabi splitting.
Near term improvements, specifically a 1.5-fold reduction in effective loss, can enable up to two orders of magnitude enhancement in sensitivity, with a projected sensitivity reaching as high as \SI{100}{\femto\tesla\per\sqrt{\hertz}}, which is very close to the Johnson-Nyquist limited magnetic noise floor of \SI{97}{\femto\tesla\per\sqrt\hertz} at ambient temperatures. 

\section{Conclusion and Outlook}

In conclusion, we have established that quantum coherence, when engineered into a structured multispectral manifold, serves as a robust resource for enhancing metrological sensitivity in solid-state sensors.
By operating a MW cavity-coupled NV ensemble in a cascaded regime, characterized by strong classical driving (Autler-Townes splitting) and spin-cavity coupling (avoided crossings), we successfully generate the coherent ``doubly dressed'' states. 
This architecture allows us to move beyond the limitations of single-mode sensing by aggregating the information distributed across the spectral peaks. Moreover, the weak cooperativity facilitates low-latency output and broadband readout. The observed sensitivity of \SI{12}{\pico\tesla\per\sqrt{\hertz}} validates the efficacy of utilizing coherent resources in cavity QED as relatively cheap and robust alternatives to the fragile multipartite entanglement.
Our theoretical model indicates that current performance is limited primarily by technical noise and thermal fluctuations, rather than fundamental constraints of the coherence resource. We predict that reducing the effective dissipation can yield sensitivities as high as \SI{100}{\femto\tesla\per\sqrt{\hertz}}. 
Crucially, this work opens a powerful possibility of using frequency multiplexing for sensitivity scaling in ambient condition quantum metrology. 
We speculate that extending our model to other hybrid architectures based on coupling of emitters to multimode resonators or a frequency comb could enable the generation of multiple coherent states. This possibility can offer a route for Heisenberg like single mode precision that leverages the robustness of coherence against the stringent demands of entanglement generation and its limitations in ambient  conditions. 

\begin{acknowledgments}
H.K. acknowledges funding from Prime Minister Research Fellowship. R.G. acknowledges fellowship support from CSIR-HRDG.
H.S.D. acknowledges financial support from SERB-DST, India via a
Core Research Grant CRG/2021/008918 and the Industrial Research \& Consultancy Centre, IIT Bombay via grant (RD/0521-IRCCSH0-001) number 2021289. K.S. acknowledges funding from Department of Science and Technology (DST) Indian National Quantum Mission (NQM) and DST SERB Power Research Grant. 
\end{acknowledgments}

\bibliography{cavity_prl}
\end{document}